\documentclass[useAMS,usenatbib]{mn2e}
\usepackage{graphicx}
\title[Young O2If*/WN6 Star in a Compact H{\sc ii} region in NGC 3603]{An O2If*/WN6 Star Catch in the Act in a Compact H{\sc ii} region in the Starburst Cluster NGC 3603}
\author[A. Roman-Lopes]{A. Roman-Lopes$^{1}$\thanks{E-mail:
roman@dfuls.cl}\\
$^{1}$Department of Physics - Universidad de La Serena - Cisternas, 1200 - La Serena - Chile\\
}

\begin{document}

\date{}

\pagerange{\pageref{firstpage}--\pageref{lastpage}} \pubyear{2010}

\maketitle

\label{firstpage}

\begin{abstract}

In this letter we report the discovery of an O2If*/WN6 star probably still partially embedded in its parental cocoon in the star-burst cluster NGC 3603.
From the observed size of the associated compact H{\sc ii} region, it was possible to derive
a probable dynamic age of no more than 600,000 years. Using the computed visual extinction value A$_V$ $\sim$ 6.0$\pm$0.2 magnitudes, an absolute visual 
magnitude M$_V$=-5.7 mag is obtained, which for the assumed heliocentric distance of 7.6 kpc results in a bolometric luminosity of 8$\times$10$^5$ L$_\odot$. 
Also from the V magnitude and the V-I color of the new star, and previous models for NGC3603's massive star population, we estimate its mass for the binary 
(O2If*/WN6 + O3If) and the single-star case (O2If*/WN6). In the former, we find that the initial mass of each component possibly exceeded 
80 M$_\odot$ and 40 M$_\odot$, while in the latter MTT 58's initial mass possibly was in excess of 100 M$_\odot$. 

\end{abstract}


\section{ Introduction}

Very massive stars (initial masses $\sim$ 100 M$_{\odot}$ or higher) are key actors in the energy balance and chemical evolution of galaxies. 
Due to their powerful winds and expanding H{\sc ii} regions, they inject large quantity of momentum and energetic ultraviolet (UV) photons
into the local interstellar medium (ISM), possibly regulating the star formation rate in its vicinity \citep{b1}.
However, one of the most fundamental yet still non-answered question in astrophysics is \textquotedblleft how do very massive stars form?\textquotedblright. 
We already have good knowledge on how 
the formation and early evolution of low mass (m$_{*}$ $\leq$ 8 M$_{\odot}$) stars occurs, but the basic processes leading to the 
formation of massive stars still remain unknown, probably because they are very rare objects
whose birthplaces are generally much more distant from us than the nearby sites of low mass star formation. Also, 
because high mass stars evolve much faster than low mass stars, they are very short lived objects, being usually deeply 
embedded into their natal environment throughout 
their very early evolutionary stages.

The very young massive star studied here was first cataloged as MTT 58 by \citet{b2}. It probably belongs to 
NGC 3603, the closest star-burst like cluster \citep{b4,b2,b5,b6} localized at an heliocentric distance of 7.6$\pm$0.4 kpc \citep{b3}. With dozens of very massive stars in its 
core, with some of them possibly presenting initial masses exceeding 100 - 150 M$_\odot$ \citep{b3}, NGC 3603 is one of the best Galactic sites for studies on the 
formation and evolution of very massive stars in the local universe.

\section{Near-Infrared spectroscopic observations and data reduction}

MTT 58 was chosen for near infrared (NIR) spectroscopic follow-up observations based on its near- to mid-infrared colors, H$\alpha$ and X-Ray
emission characteristics (the details on the criteria and general selection methodology are fully discussed in a forthcoming paper - Roman-Lopes submitted).
The NIR spectroscopic observations were performed with the Ohio State Infrared Imager and Spectrometer (OSIRIS)
at the Southern Astrophysics Research (SOAR) telescope. The J-, H- and K-band data were acquired in 9th May 2011 with the night presenting good seeing 
conditions. Besides
MTT 58, we also obtained NIR spectra for HD93129A (O2If*), WR20a (O3If*/WN6 + O3If*/WN6) and WR42e (O2If*/WN6) \citep{b7,b8}. 
In Table 1 it is shown a summary of the NIR observations used in this work. 

   \begin{figure}
    \hspace{-10pt}
   \centering
   \includegraphics[bb=14 14 401 293,width=8.2 cm,clip]{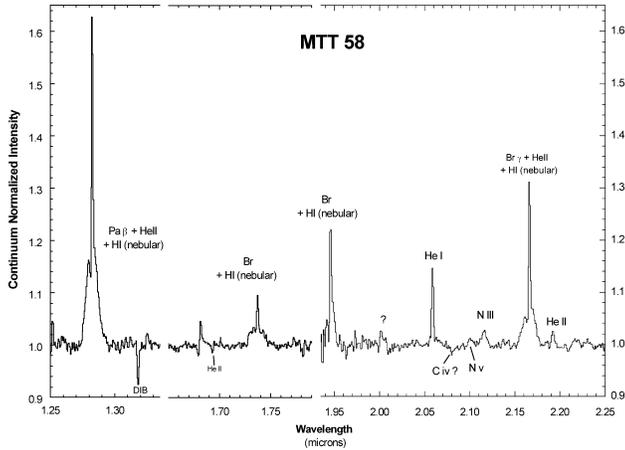}
      \caption{The J- H- and K-band continuum normalized SOAR-OSIRIS spectra of MTT 58, with the main H, He and N emission lines identified by labels. 
      Notice the intense nebular hydrogen recombination lines superimposing the broad emission lines produced by the powerful wind of the embedded star.}
         \label{FigVibStab}
   \end{figure}

The raw frames were reduced following standard NIR reduction procedures. 
The two-dimensional frames were subtracted for each pair of images taken
at the two shifted positions. Next the resultant images were divided
by a master normalized flat, and for each processed frame, the J-, H- and K-band spectra were extracted using the IRAF task 
APALL, with subsequent wavelength calibration being performed using the IRAF tasks IDENTIFY/DISPCOR
 applied to a set of OH sky line spectra (each with about 30-35 sky
lines in the range 12400\AA\ -23000\AA\ ). 
The typical error (1-$\sigma$) for this calibration process is estimated as
$\sim$12\AA\, which corresponds to half of the mean FWHM of the OH lines in
the mentioned spectral range. 
Telluric atmospheric corrections were done using J-, H- and K-band spectra of A
type stars obtained before and after the target observations. 
The photospheric absorption lines present in the high signal-to-noise telluric
spectra, were subtracted from a careful fitting (through the use of Voigt and
Lorentz profiles) to the hydrogen absorption lines and respective adjacent continuum. 
Finally, the individual J-, H- and K-band spectra were combined by the average (using the IRAF task SCOMBINE) with the mean signal-to noise ratio of the resulting spectra
well above 100.


\begin{table}
\caption{Summary of the SOAR/OSIRIS dataset used in this work.}
\label{catalog}
\centering
\renewcommand{\footnoterule}{}  
\begin{tabular}{cc}
\hline
   Date  & 09/05/2011\\
   Telescope  & SOAR\\
   Instrument & OSIRIS\\
   Mode  & XD\\
   Camera & f/3\\
   Slit  & 1" x 27"\\
   Resolution  & 1000\\
   Coverage ($\mu$m) & 1.25-2.35\\
   Seeing (")  & 1-1.5\\
\hline
\end{tabular}
\end{table} 


   \begin{figure}
    \hspace{-10pt}
   \centering
   \includegraphics[bb=14 14 681 520,width=8 cm,clip]{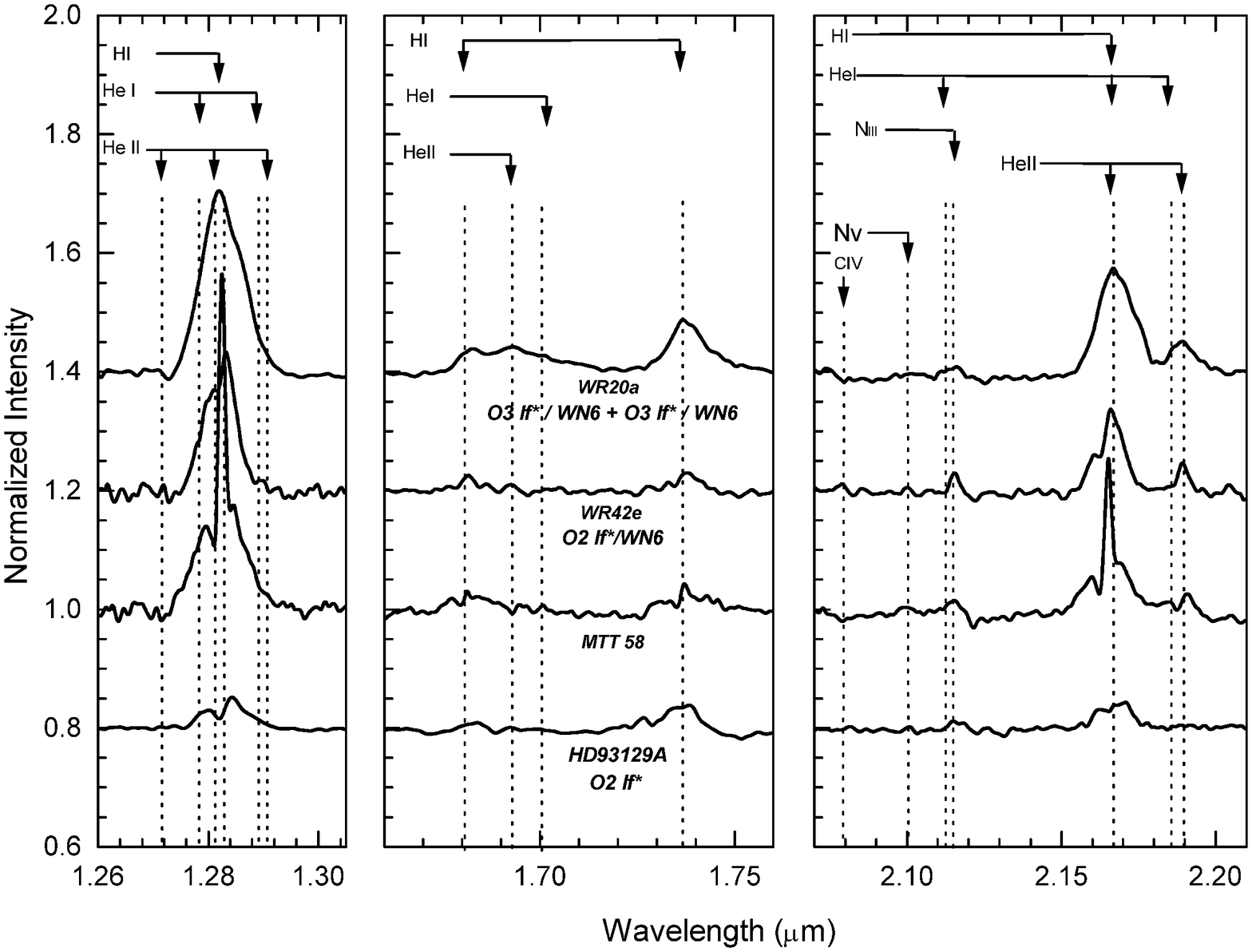}
      \caption{The J- H- and K-band continuum normalized SOAR-OSIRIS spectra of the MTT 58, together with the NIR spectrograms of
               HD93129A (O2If*), WR42e (O2If*/WN6) and WR20a (O3If*/WN6 + O3If*/WN6), with the main H, He and N emission lines identified by labels. Notice the strong nebular
               hydrogen recombination narrow lines superimposing the broad emission lines produced by the powerful wind of the embedded star.
               Besides the nebular emission components, the MTT 58's J-, H- and K-band spectra resemble well those of WR42e.}
         \label{FigVibStab}
   \end{figure} 

   \begin{figure*}
  \centering
  \includegraphics[bb=14 14 420 447,width=10 cm,clip]{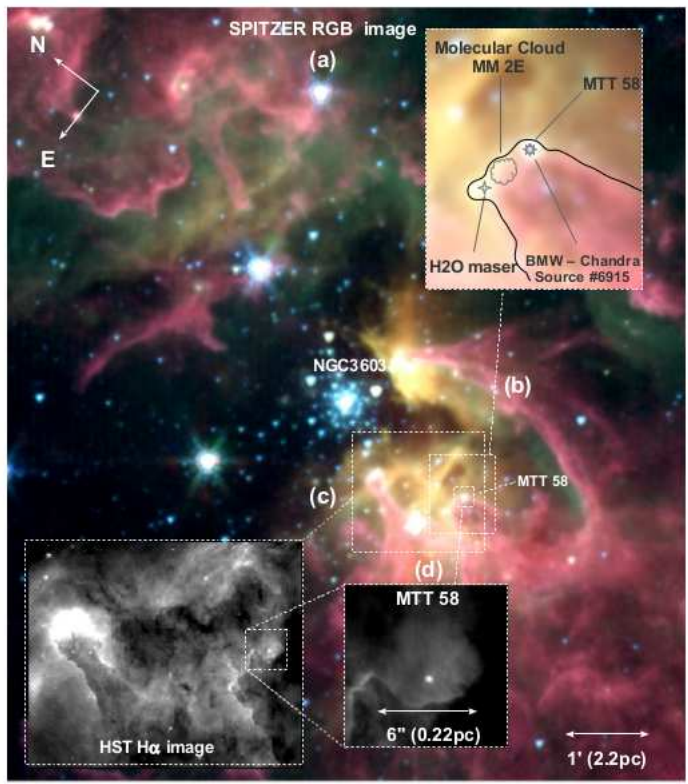}
     \caption{(a) The Spitzer false-color RGB image of the region centered on the core of the NGC3603 star-burst cluster. (b) Zooming of the region in the direction of
     MTT 58, with the associated pillar (delineated by the black line) mostly visible in the red channel. There we also can see the positions of the BMW - Chandra source 
     \#6915 \citep{b13}, the molecular cloud MM 2E \citep{b14} and the H$_2$O maser
     found by \citet{b15}. (c) The HST H$\alpha$ image of NGC 3603 where we can see the spectacular pillars to the south of the cluster core). 
     (d) Detailed view of the H$\alpha$ emission in the vicinity of MTT 58, where we can see the H$\alpha$ counterpart of part of the compact H{\sc ii} region detected 
     by the ATCA observations.}
        \label{FigVibStab}
   \end{figure*}

\section{Results}

Coordinates and photometry of MTT 58 are shown in Table 2. The B-, V- and I-band magnitudes were taken from
the work of \citet{b20}, while the NIR values were obtained from the Two-Micron All Sky Survey \citep{b21}, with the absorption-corrected 0.5-10keV
Chandra X-ray flux taken from the work of \citet{b13}.

\subsection{The OSIRIS NIR spectra of MTT 58: An O2If*/WN6 star embedded in a compact H{\sc ii} region}

\begin{table*}
\begin{center}
\caption{Coordinates (J2000), Optical/NIR photometry, and X-ray parameters of the newly-identified O2\,If*/WN6 star.
The BVI photometry was taken from \citet{b20}, while the near-infrared magnitudes are from \citet{b21}. Finally, the absorption-corrected 0.5-10keV flux is from the 
work of \citet{b13}.\label{tbl-2}}
\begin{tabular}{ccc}
\\
\hline\hline
 RA=11h15m07.60s   &  Dec=-61d16m54.8s &     \\
\hline
B = 16.14  & V = 14.76 & \\
\hline
I = 12.39 & J = 10.47$\pm{0.02}$ &  \\
\hline
H = 9.68$\pm{0.02}$ & K$_S$ = 9.24$\pm{0.02}$ \\
\hline
\centering X-Ray (0.5-10 keV) = 4.23$\times10^{-17}$Wm$^{-2}$   &   \\
\hline\hline

\end{tabular}
\end{center}
\end{table*}

Figure 1 shows the telluric corrected (continuum normalized) J-, H- and
K-band SOAR-OSIRIS spectra of MTT 58. They present (despite of the previous subtraction of the background extended nebular components) strong residual hydrogen 
recombination features 
that are particularly prominent in the Pa$\beta$ and Br$\gamma$ transition lines. They appear as very narrow lines superimposing the broad emission lines generated by 
the strong stellar wind of the embedded star. The presence of strong narrow nebular line emission indicates that the star is probably immersed in an ionized region. 

   \begin{figure*}
   \centering
  \includegraphics[bb=14 14 668 637,width=9 cm,clip]{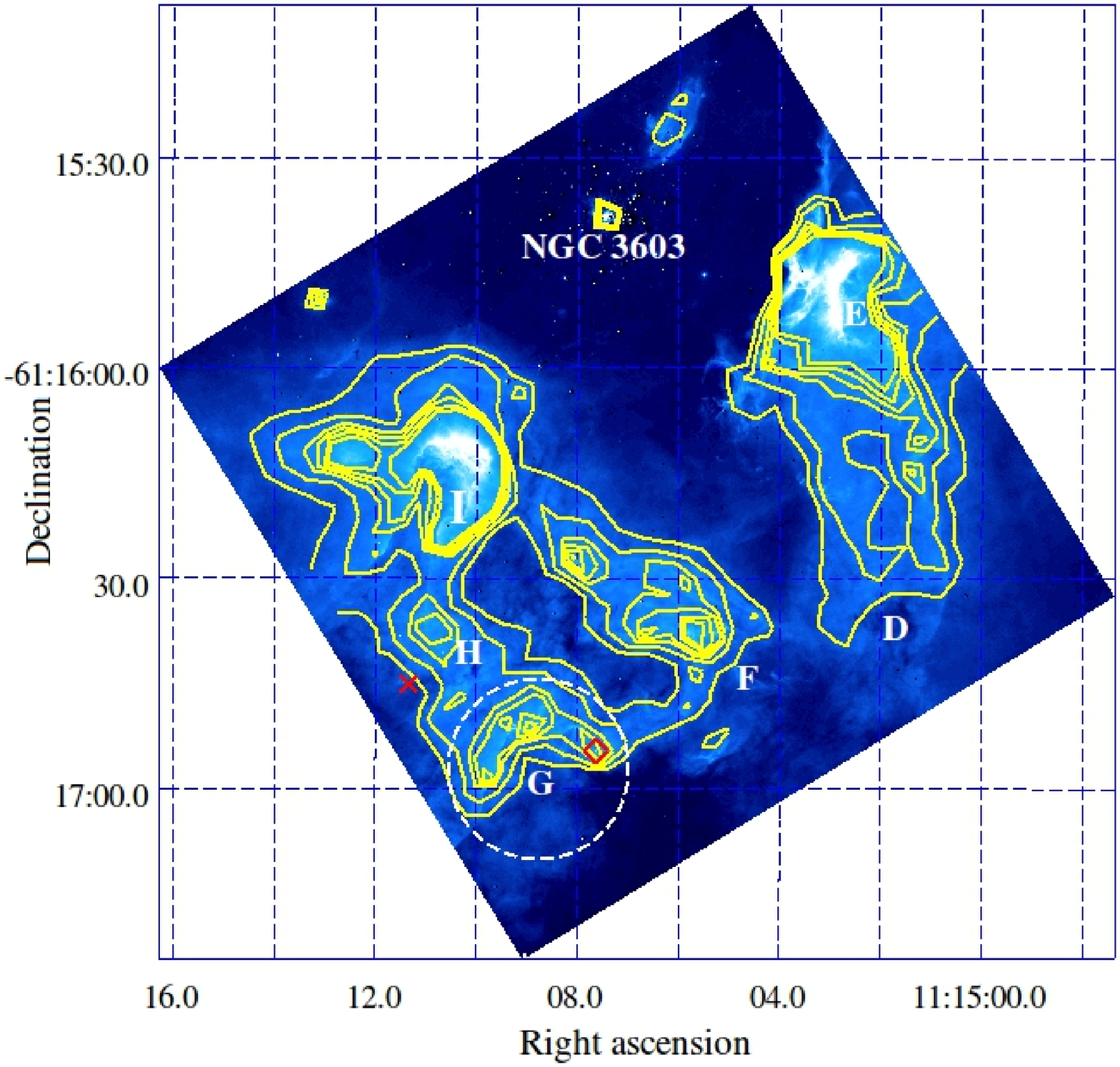}
  \caption{The continuum subtracted HST H{$\alpha$} image (logarithmic scale) of the region to the south of NGC 3603 cluster. 
  For the sake of clarity, we also 
  present the H{$\alpha$ intensity contours (yellow lines) in steps of 4500, 5500, 6500, 7500 and 8500 units (arbitrary scale).}
  There we indicate the location of the NGC 3603's cluster and the position of the
  compact H{\sc ii} regions D, E, F, G, H and I detected by \citet{b10} with the Australia Telescope Compact Array (ATCA). The estimate angular size of 
  the compact H{\sc ii} 
  region G, is indicated by the white dashed circle. Also, the location of the O2If*/WN6 star is indicated by the red diamond. Notice the extended emission in the vicinity 
  of MTT 58.}
         \label{FigVibStab}
   \end{figure*}

Figure 2 shows the MTT 58's J-, H- and
K-band SOAR-OSIRIS spectra, together with those for HD93129A (O2If*), WR42e (O2If*/WN6) 
and WR20a (O3If*/WN6 + O3If*/WN6).
The MTT 58's J-, H- and K-band spectra resembles well (besides the nebular emission line components) those of WR42e, indicating that the star is a 
new exemplar of the O2If*/WN6 type. 
Such objects are rare members of the WNH group that is probably compound by the most massive hydrogen core-burning stars known, which due to their high intrinsic 
luminosities (close to the Eddington limit), 
possess emission-line spectra that mimic the spectral appearance of classical WR stars, even in the beginning of their life times \citep{b9}.
As a last comment on the MTT 58's spectra, it is interesting to notice that
there are two identified lines that appears in absorption. The first is a line at 1.693$\mu$m, that could be due to He{\sc ii}, while the other is the
C{\sc iv} line at $\sim$ 2.080$\mu$m. Such absorption lines could be indicative of the presence of an early-O star companion.
Indeed, the presence of an X-ray source gives support for this idea.

Figure 3(a) shows a false color RGB image made from the 3.6$\mu$m (blue), 4.5$\mu$m (green) and 5.8$\mu$m (red) Spitzer IRAC images of the region centered in the core 
of the NGC 3603. 
The bulk of the remnant of the NGC 3603's parental molecular gas cloud is clearly seen to the south and south-west, where the large-scale star formation 
is still taking place \citep{b10,b12,b11,b6}. As we can see from the figure, MTT 58 is placed at about 1.8 arcmin to the south of the NGC 3603's centre, at the tip 
of a giant pillar of gas and dust. Figure 3(b) presents a detailed view of the region around MTT 58. There we indicate (besides the star) the position 
of the X-Ray BMW-Chandra point source \#6915 \citep{b13}, and that for the molecular cloud MM 2E \citep{b14} and the H$_2$O maser found by \citet{b15}. 
They appear very close in projection (0.15 pc and 0.33 pc, respectively) to MTT 58, 
with the X-ray source coordinates coinciding exactly with the position of the star. 
The proximity to a non-destroyed molecular cloud and H$_2$O maser source, suggest that MTT 58 had no time to completely dissipate its 
parental molecular cocoon, indicating that it is probably an extremely young O2If*/WN6 star. 

\subsection{Size and age of the MTT 58's H{\sc ii} region}

As mentioned in the previous section, MTT 58 is possibly powering a compact H{\sc ii} region.
We searched in the literature for high spatial resolution radio continuum observations that could
give us constrains on the size of the MTT 58's H{\sc ii} region, and found that \citet{b10} observed the region towards MTT 58 with the Australia 
Telescope Compact Array (ATCA), 
in the continuum (at 8.8 GHz) and the H90$\alpha$, He90$\alpha$, C90$\alpha$ and H113$\beta$ recombination 
lines (all with rest frequencies around 8.9 GHz), with angular resolution of 7". Their source G 
with angular size of 26" and integrated flux density S$_{tot}$=4.4$\pm$0.3 Jy, is one of the strongest sources in the entire region. It
has coordinates $\alpha$=11:15:08.76 and $\delta$=-61:16:55.7 (J2000) that matches (considering the associated uncertainties) those of MTT 58. 
Taking into account the limited (7") ATCA's spatial resolution, we may 
consider 26" as an \textit{upper limit} for the angular size of the MTT 58's compact H{\sc ii} region.

We also searched in the 
Hubble Legacy Archive\footnote{http://hla.stsci.edu/hlaview.html} looking for H$\alpha$ images of NGC 3603, founding that the region towards MTT 58 was observed 
in the framework of the proposal ID \#11360 (P.I. O'Connell, R. W.).
Figure 3(c) shows the 
HST H$\alpha$ image of the southeast part of the NGC 3603's region, where we can see the bright H$\alpha$ extended emission generated 
in the top of a spectacular pillar visible to the north of MTT 58. Figure 3(d) presents a detailed view of the H$\alpha$ emission in the vicinity of MTT 58. 
By scaling the F658N ([N{\sc ii}] 6583) image to the F656N (H$\alpha$ 6552) one
(using a set of non-saturated stars), and subtracting the former from the last we obtained a 
continuum subtracted HST H{$\alpha$} image of the region to the south of NGC 3603 cluster. The resulting image is shown in Figure 4 with 
the location 
of the NGC 3603 cluster, MTT 58 and the ATCA compact H{\sc ii} regions D, E, F, G, H and I \citep{b10} indicated by labels.
As a complement and in order to compare the H{$\alpha$} subtracted image with the positions of the 8.8 GHz compact continuum sources detected by 
\citet{b10}, we also present the H{$\alpha$} intensity 
contours (represented by the yellow continuum lines), which correspond to 4500, 5500, 6500, 7500 and 8500 counts (arbitrary scale).  
The H{$\alpha$} contours associated to the ATCA source G, present an arc shaped structure (not spherical like in the radio continuum map of \citet{b10}, 
probably due to the presence of very dense 
foreground molecular material well seen (against the bright background extended emission) in Figure 3(b). 

We can now estimate the age of the 
embedded star from the H{\sc ii} region's size, assuming that it reached its initial Str\"omgren
radius in a very short time (a few 10$^4$ yrs), corresponding to a rapid
expansion phase dominated by an R-type shock \citep{b16},
and that after this phase it has been expanding in a uniform medium
owing to the pressure difference between the hot ionized gas and
the outer cool molecular gas.
This corresponds to a condition where
the H{\sc ii} region expansion is governed by a weak D-type ionization
front, with \textit{a rate of expansion} that can be estimated using the
equation \citep{b16}:


R$_f$(t) = R$_i$ (1+$\frac{7C{\sc II}}{4R_i}$t)$^{\frac{4}{7}}$                


where R$_i$ and R$_f$ are the initial and final values for the Str\"omgren radius,
and \textit{CII} is the sound speed in the H{\sc ii} region (typically varying from 0.4 to 
10 km/s \citep{b29,b25,b30,b27,b28,b26}. In order to be conservative, in the
calculation we will assume a mean sound speed of 4 km s$^-$$^1$ and
that the numerical density in the beginning
of the ultra-compact phase was about 10$^5$ cm$^-$$^3$ \citep{b18}, and that the number of Lyman continuum photons (N$_{Ly}$) emitted by second by an O2If*/WN6 
star right in the very
beginning of its life-time is about 10$^{49}$ s$^-$$^1$. This can be considered a lower limit because the earliest O super-giant stars are believed to emit  
much more than 10$^{49}$ s$^-$$^1$ Lyman continuum photons during its main sequence life-time \citep{b17}.

   \begin{figure*}
   \centering
  \includegraphics[bb=14 14 325 433,width=6 cm,clip]{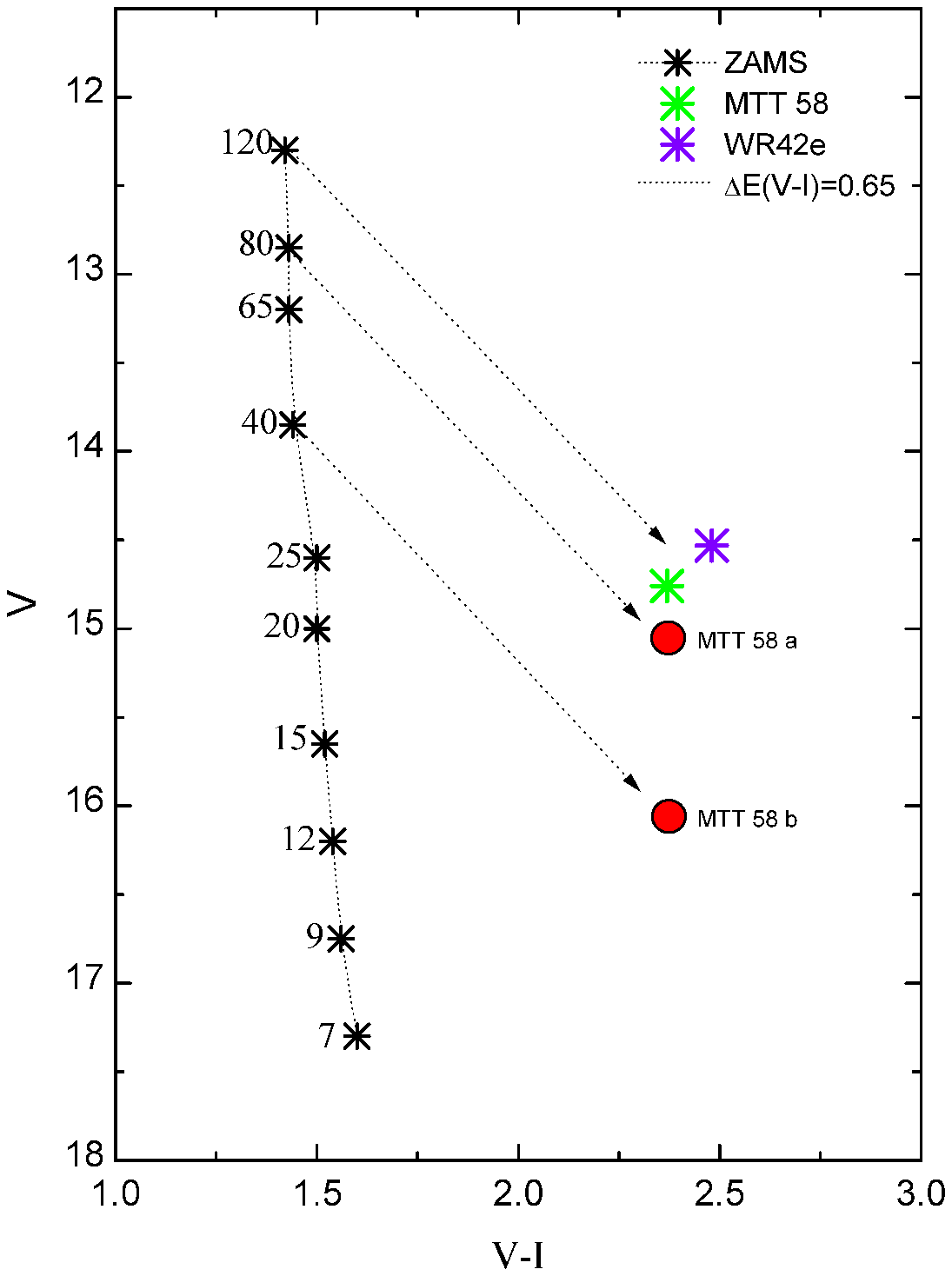}
  \caption{The V $\times$ (V-I) diagram (based on Figure 7 of the work of \citet{b20}, with MTT 58 represented by the green star
  (single star case) and red dots (\textbf{binary} system case), and WR42e $\sim$ 130M$_\odot$ - \citep{b24} by the violet star.
  The non-reddened main sequence at the quoted distance of 7.6 kpc \citep{b3} for masses between 7M$_\odot$ to 120M$_\odot$ is represented by the black dotted line,
  with the black stars indicating the position of each mass bin. Also, 
  the reddening vector taken from the work of \citet{b20} is represented by the line-dotted arrows. From this diagram we can see that the initial mass of the 
  new O2If*/WN6 star (in the single star case) possibly exceed 100M$_\odot$. On the other hand, if MTT 58 is assumed to be a binary system compound by
  a O2If*/WN6 + O3If* system, the individual masses should exceed 80$_\odot$ and 40$_\odot$, respectively.}
         \label{FigVibStab}
   \end{figure*}


From the angular size of the source G obtained by \citet{b10}, we can estimate its corresponding dynamical age.
The Str\"omgren radius \citep{b23} for an initial numerical density of 10$^5$ cm$^-$$^3$ and N$_{Ly}$=10$^{49}$ s$^-$$^1$ is R$_i$ $\sim$ 0.025 pc. 
The upper limit for the final radius R$_f$ can be computed from the angular size (26") estimated from the ATCA observations. Assuming a distance of 7.6 kpc, the 
corresponding value is R$_f$ = 0.48 pc, which applied in Equation 1 results in a dynamical age of $\sim$580,000 years. 
On the other hand, in the case in which MTT 58's N$_{Ly}$=10$^{50}$ s$^-$$^1$ \citep{b7}, the dynamical age would drop by a factor $\sim$3.8, 
resulting in a much lower age of
150,000 yrs! In any scenario, we conclude that MTT 58 is possibly the youngest Galactic O2If*/WN6 star found to date.

Recently, \citet{b8} reported the discovery of an O2If*/WN6 star (WR42e) that is thought to have been ejected from the NGC 3603 cluster core. In this sense
we may argue that in the MTT 58's case the situation is probably different. Indeed, considering that an O2If*/WN6 star 
produces much more Lyman continuum photons than the template (O7{\sc v} star with Q$_0$ $\sim$10$^{49}$ s$^-$$^1$) we used in our previous calculations, in this case the 
tremendous ionization front
of the O2If*/WN6 star could propagate much quicker than the associated star travel velocity, destroying and dissociating any molecular cloud in its way much earlier of 
its arrive there. On the other hand, another argument favoring the idea that we are probably looking at an in-situ formation case, is the fact that the observed 
morphology of the compact H{\sc ii} region studied by \citet{b10} is not cometary, like the expected when the ionizing source is traveling with a relatively high 
transverse velocity \citep{b18}.

\subsection{Estimating the luminosity of MTT 58}

In order to compute estimates for the luminosity and mass of MTT 58, we need to evaluate its visual extinction taking into account that 
the interstellar reddening law for NGC 3603 is probably abnormal \citep{b19,b20}, with a ratio of total to selective extinction value R$_V$=3.55$\pm$0.12 \citep{b20}.
From Table 2, we can see that
MTT 58 presents (B-V) color $\sim$ 1.4 mag, which for an assumed mean intrinsic (B-V)$_0$ value of -0.3 mag (typical for the hottest early-type stars), corresponds 
to a color excess E(B-V) $\sim$ 1.7 mag (A$_V$ $\sim$ 6.0$\pm$0.2 mag), a extinction value higher than that derived by \citet{b20} for the NGC 3603 early-type members 
(E(B-V)=1.25, A$_V$=4.4$\pm$0.1), but still 
compatible with what is expected for an early-type member of NGC 3603 \citep{b20}. Assuming that the additional amount of reddening A$_V$=1.6 is generated by the gas 
and dust present in the compact HII region (e.g. local), we can speculate that the embedded star could be close to disrupt the border of the cavity that is facing 
the MTT 58's line of sight.

From the computed color excess, one can estimate the MTT 58's absolute magnitude using the distance modulus equation, 
assuming that the star is placed at an heliocentric distance of 7.6$\pm$0.4 kpc \citep{b3}. We computed M$_V$=-5.7 mag (or M$_K$=-5.9 
considering A$_{K_s}$=0.12A$_V$ - \citet{b3}, which are values that are a bit lower that those obtained by other researchers for stars of similar type \citep{b3,b22,b7}.
However, it is also known that massive stars at their very beginning stages are expected to be less luminous then similar stars at the more evolved main-sequence
phase.
Considering the derived absolute visual magnitude and assuming a mean bolometric correction BC $\sim$ -4.3 mag \citep{b3,b7}, we estimate the bolometric magnitude of
MTT 58 as M$_{Bol}$ $\sim$ -10.0, which corresponds to a total stellar luminosity above 8$\times$10$^{5}$ L$_\odot$.

\subsection{Binarity}

In Section 3.1 it was mentioned that the H-band MTT 58's spectrum shows the He{\sc ii} line at 1.693$\mu$m in absorption, which could indicate the presence of 
an early-O star companion. As MTT 58 has an associated X-ray source, it is useful to compare its bolometric and X-ray luminosities. 
From its absorption-corrected 0.5-10keV flux (Table 2), and using the adopted heliocentric distance of 7.6$\pm$0.4 kpc \citep{b3} 
we compute an X-ray luminosity L$_X$ $\sim$ 2.9$\times$10$^{32}$ erg s$^{-1}$ that compared with the bolometric luminosity derived in Section 3.3 results in 
L$_X$/L$_{Bol}$ $\sim$ 10$^{-6}$. This result is about ten times greather than the canonical value expected for single stars, e.g. 
L$_X$/L$_{Bol}$ $\sim$ 10$^{-7}$ \citep{b31}, favoring the idea that the O2If*/WN6 star probably has an early-O star companion.

If we assume that MTT 58 is a binary star, and from an inspection of their NIR combined
spectra (Figure 1), in principle we may conclude that the absence of any He{\sc i} line in absorption (e.g. at 1.701$\mu$m or at 2.113$\mu$m) indicates 
that the companion of the O2If*/WN6 star 
should be of spectral type earlier than O4 \citep{b32}. Indeed, from their NIR spectra of O and early-B stars, one can see that the H-band 
spectra of O3 stars (like HD 64568 (O3 V) and Cyg OB2 \#7 (O3 If*) present absorption He{\sc ii} lines at 1.693$\mu$m that are not as strong as those
found in stars later than O4. On the other hand, the 2.189 He{\sc ii} lines are 
relatively strong in the K-band spectra of class V stars, becoming more complex in absorption and in emission in the K-band spectrum the O3 If* stars. 
In this sense, the absence of a strong signature (e.g. in absorption) of the He{\sc ii} line at 2.189
in the MTT 58's K-band spectrum, enable us (in principle) to rule out a class V type companion star. 
As mentioned in Section 3.1 the MTT 58's K-band spectrum shows an absorption line at $\sim$2.080 microns, which from 
the associated wavelength we had tentatively identified as due to a C{\sc iv} line. Interestingly, the K-band spectrum of the 
O3If* Cyg OB2 \#7 does show (Hanson et al. 2005) such \textbf{an} absorption line!

We searched in the HST archive looking for high
quality images of the region towards MTT 58 (besides the H$\alpha$ image presented in Figure 3).
From a careful inspection of the point spread function of MTT 58 in the HST archive images, we did not find any evidence of a companion, indicating that if MTT 58 is 
a binary, it is probably formed by a very close system.
On the other hand, another possible explanation is that the brightness difference between the two stars is too large (by several magnitudes). In this case, 
it would be very difficult (if not impossible) to properly separate the brightness distribution of the 
secondary from that of the main component. However, if the secondary star is assumed to be an O3If* (as the spectral information suggests), 
the difference in brightness for an O2If*/WN6 star should 
not be that high. Indeed, from the work of \citet{b7} one finds that the brightness difference (in terms of bolometric magnitudes) for such kind of stars is possibly 
not greater than $\sim$ 1-1.5 magnitudes.
Further NIR and optical spectrophotometric monitoring are certainly necessary to obtain more clues for our assumption of an
O2If*/WN6 + O3If* binary system. In this sense, our group in La Serena is planning a long term spectrophotometric monitoring study of this system.

\subsection{Mass}

We can estimate the mass of MTT 58 by comparing its observed V magnitude and V-I color 
with those of other NGC 3603 cluster members, presented in Figure 7 of the work of \citet{b20}. We do that taking into account both, the scenario where it is 
assumed to be a binary system, as well as that where it is considered as a single star.
Also in order to be conservative, we assume that the difference in magnitudes between two stars of O3If* and O2If*/WN6 types is about 1 magnitude \citep{b7}, and to
simplify the process we assume that both stars have approximately the same bolometric corrections (a reasonable assumption considering their probable very
early spectral types).

Figure 5 shows an adapted version of the V $\times$ (V-I) diagram for NGC 3603 \citep{b20}, with MTT 58 (single and binary cases) and WR42e (estimated mass 
of $\sim$ 130 M$_\odot$ - \citet{b8,b24} represented by a green star (single star case) and red dots (binary system case) and violet star, 
respectively. We added WR42e in the mentioned diagram for comparison purpose
because it is of the same spectral type of MTT 58 and probably belongs to the same complex.
From this diagram we can see that for the scenario where MTT 58 is considered to be a binary system, the masses of each component would be 
about 80 M$_\odot$ (for MTT 58a) and 40 M$_\odot$ (for MTT 58b). 
This numbers can be considered reasonable if compared with those from two known very massive binary systems (with similar spectral type and morphology) 
WR20a (O3If*/WN6 + O3If*/WN6), with masses of 83 and 82
M$_\odot$ \citep{b33,b9}, and WR21a (O3If*/WN6 + early-O), for which \citep{b34} estimated minimum masses of 87 and 53 M$_\odot$, 
respectively.
Finally, in the case in which MTT 58 is assumed to be a single star, its initial mass could exceed 100 M$_\odot$.

\section{Summary}

In this work we report the discovery of an O2If*/WN6 star probably still embedded in its parental cocoon in the star-burst cluster NGC 3603.
The new O2If*/WN6 star was previously cataloged as MTT 58 by \citet{b20}, being apparently placed at the tip of a giant pillar of gas and dust at 
about 1.8 arcmin ($\sim$ 4 pc for the quoted distance of 7.6 kpc - \citep{b3} to the south of the NGC 3603's cluster center.
The proximity to a non-destroyed molecular cloud, suggests that MTT 58 yet had no time to completely dissipate its 
parental molecular cocoon, indicating that it is probably an extremely young O2If*/WN6 star.

Another interesting result is that the new O2If*/WN6 star may be the main component of a binary system. Indeed, the presence of a 
BMW-Chandra X-ray point source coincident with the MTT 58's coordinates and the fact that its NIR spectra present two absorption 
lines (the He{\sc ii} 1.693$\mu$m and the 2.080$\mu$m C{\sc iv} lines) possibly generated by an O3If* companion, 
give strong support for this idea.

From the observed size of the associate compact H{\sc ii} region detected at 3.4 cm by \citet{b10} using the Australian Telescope Compact Array (ATCA), 
it was possible to derive
a probable dynamic age of no more than 600,000 years. From the computed visual extinction value A$_V$ $\sim$ 6.0$\pm$0.2 mag an absolute visual 
magnitude M$_V$=-5.7 mag is obtained, which for the assumed heliocentric distance of 7.6 kpc results in a bolometric luminosity of 8$\times$10$^5$ L$_\odot$. 

Finally, from the V magnitude and V-I colour of the new O2If*/WN6 star and the Figure 7 of the work of \citet{b20}, we estimate the MTT 58's mass considering
the cases in which it is assumed to be a binary system (O2If*/WN6 + O3If stars), and the one in which it is thought to be a single O2If*/WN6 star.
In the first case we found that the initial masses of MTT 58a and MTT 58b should be 
above 80 M$_\odot$ and 40 M$_\odot$ respectively. On the other hand, in the scenario were MTT 58 is assumed to be a single star, its initial mass possibly 
exceeded 100 M$_\odot$.

\section*{Acknowledgments}

This research has made use of the NASA/ IPAC Infrared Science Archive, which
is operated by the Jet Propulsion Laboratory, California Institute of
Technology, under contract with the National Aeronautics and Space
Administration. 
This publication makes use of data products from the Two Micron All Sky
Survey, which is a joint project of the University of Massachusetts and the
Infrared Processing and Analysis Center/California Institute of Technology,
funded by the National Aeronautics and Space Administration and the National
Science Foundation. 
Based on observations obtained at the Southern Astrophysical Research (SOAR) telescope, which is a joint project of the 
Minist\'{e}rio da Ci\^{e}ncia, Tecnologia, e Inova\c{c}\~{a}o (MCTI) da Rep\'{u}blica Federativa do Brasil, the U.S. National 
Optical Astronomy Observatory (NOAO), the University of North Carolina at Chapel Hill (UNC), and Michigan State University (MSU).
Also, this research has made use of the SIMBAD database, operated at CDS,
Strasbourg, France. 
This work was partially supported by the Department of Physics of the Universidad de La Serena.
ARL thanks financial support from Diretoria de Investigaci\'on - Universidad de La Serena through Project \textquotedblleft 
Convenio de desempe\~no DIULS CDI12\textquotedblright.



\end{document}